\documentclass[review,12pt,numafflabel]{elsarticle}    
\usepackage{figure/sty_thang_elsarticle}
\usepackage[sfdefault]{roboto}
\usepackage{pdfpages}

\begin{document}
\begin{frontmatter}
  \title{CLFF: Automated Framework for Concurrent Learning Force Fields and Material Property Calculations}

  \author[au1]{Cao Thang Nguyen}
  \author[au1]{Sung Youb Kim \corref{cor1}}
  \cortext[cor1]{Corresponding Author: sykim@unist.ac.kr}

  \affiliation[au1]{organization={Department of Mechanical Engineering, Ulsan National Institute of Science and Technology}, city={Ulsan}, postcode={44919}, country={South Korea}}

  \begin{abstract}
    Machine learning interatomic potentials (MLIPs) offer an efficient paradigm for large-scale atomistic simulations with near first-principles accuracy. However, developing and deploying these potentials often involves intricate procedures that require substantial manual effort and domain expertise. In this work, we present CLFF, an open-source Python framework designed to automate concurrent learning workflows for MLIPs. CLFF integrates model training, MLIP-driven sampling, uncertainty-based candidate selection, quantum-mechanical labeling, and dataset augmentation into a unified workflow. The framework currently supports several modern graph-based MLIP architectures and is compatible with widely used computational chemistry and materials science tools. A distinguishing feature of CLFF is its ability to asynchronously distribute computational tasks across multiple heterogeneous remote computing resources, enabling users to aggregate fragmented hardware infrastructure within a single controllable workflow. In addition to concurrent learning, CLFF provides automated workflows for initial data generation and material property calculations, including phonon dispersion and potential energy surface scans. By minimizing manual intervention and enhancing reproducibility, CLFF offers a flexible platform for developing and deploying MLIPs in computational materials research.
  \end{abstract}

\end{frontmatter}

\section{Introduction}
Molecular Dynamics (MD) simulations have become indispensable tools in materials science, providing atomic-level insights into complex material behaviors. Although \textit{ab initio} molecular dynamics (AIMD) yields reliable atomic forces by explicitly treating electronic structures, its high computational cost restricts simulations to small systems and short timescales \cite{parr1995DensityFunctional}. Conversely, classical MD using empirical potentials is several orders of magnitude faster, making it highly attractive for large-scale simulations \cite{frenkel2002Understanding}. However, the predictive power of classical MD depends entirely on the quality of the underlying interatomic potential (IP), and minimizing the trade-off between accuracy and efficiency remains a long-standing challenge.

Recent advancements in machine learning (ML) have revolutionized materials modeling through Machine Learning Interatomic Potentials (MLIPs). MLIPs bridge the gap between empirical efficiency and first-principles accuracy, enabling large-scale simulations at near \textit{ab initio} resolution \cite{mortazavi2025Recent, jia2020pushing}. Developing an MLIP requires a trustworthy quantum-mechanical dataset and an efficient architecture. The latter has been extensively explored, leading to a diverse array of MLIP architectures, ranging from kernel-based models \cite{bartok2010Gaussian, thompson2015Spectral, drautz2019Atomic} and classical neural networks \cite{behler2007Generalized, smith2017ANI1, zhang2018Deepa} to highly efficient, modern graph-based architectures \cite{batzner2022E3equivariant, batatia2023MACE, park2024Scalable, liao2026EquiformerV3}. In contrast, the curation of trustworthy datasets has received comparatively less attention. Public Density Functional Theory (DFT) datasets cover various systems \cite{hoja2021QM7X, ramakrishnan2014Quantum, levine2026Open, haastrup2018Computational, jain2013Commentary, gharakhanyan2025Open, barroso-luque2026Open, ramlaoui2025LeMatTraj} but often exhibit varying accuracy levels due to disparate configuration parameters (e.g., exchange-correlation functionals, energy cutoffs, convergence criteria, and van der Waals corrections) and are frequently insufficient for application-specific contexts. Dataset quality and diversity fundamentally govern MLIP performance. A dataset must represent the configurational space encountered during simulations; insufficient coverage results in unphysical predictions, while redundant data causes overfitting. Curating a representative dataset manually is challenging, susceptible to user bias, and requires fragmented execution across complex computational toolchains.

To automate MLIP dataset curation, concurrent learning schemes have attracted considerable interest \cite{podryabinkin2017Active, jinnouchi2020OntheFly, zhang2020DPGEN}. Distinct from traditional active learning, concurrent learning in the context of MLIPs involves dynamically training the model, exploring configurational space, and curating the dataset on the fly. A typical concurrent learning cycle involves: (1) training initial MLIPs on a small baseline dataset, (2) exploring the configurational space using inexpensive MLIP-driven sampling (e.g., MD or Enhanced Sampling) to identify highly informative configurations, and (3) labeling these configurations using high-accuracy calculations (e.g., DFT) to augment the training dataset. This cycle iterates until the MLIP achieves satisfactory accuracy. While several automated frameworks employ this strategy to accelerate MLIP development \cite{bidoggia2026AiiDATrainsPot, verma2026SPARC, chen2025NepTrain, novikov2020MLIP, zhang2020DPGEN, schutt2019SchNetPack}, most are restricted to specific architectures or sampling techniques, or lack comprehensive integration with high-performance computing (HPC) schedulers. For instance, DP-GEN \cite{zhang2020DPGEN} effectively manages computational resources but is tailored primarily for DeepMD-based models \cite{zeng2023DeePMDkit, zeng2025dpdataa}, lacking direct compatibility with contemporary graph neural networks. Furthermore, existing packages typically lack the capability to distribute workloads across heterogeneous resources on multiple remote machines, limiting scalability and efficiency.

This study introduces CLFF (Concurrent Learning Force Field), an open-source framework featuring multiple end-to-end automated workflows for the concurrent learning of MLIPs and material property calculations. CLFF streamlines the pipeline by integrating cutting-edge algorithms into a unified, generalized workflow. The framework features broad compatibility with existing software: it natively supports various graph-based MLIP architectures (e.g., SevenNet \cite{park2024Scalable}, MACE \cite{batatia2023MACE}, NequIP \cite{batzner2022NequIP}, TensorNet \cite{simeon2023TensorNet}) with an extensible design; integrates seamlessly with LAMMPS \cite{thompson2022LAMMPS}, ASE \cite{hjorthlarsen2017atomic}, and GPAW \cite{mortensen2024GPAW}; supports enhanced sampling methods via PLUMED \cite{tribello2014PLUMED}; includes van der Waals (vdW) corrections \cite{grimme2010consistent, grimme2011Effect, smith2016Revised}; and utilizes the DPDispatcher package \cite{yuan2025DPDispatcher} for remote connectivity. Additionally, CLFF implements complementary workflows for initial dataset generation and automated property calculations. A core capability of CLFF is its native support for asynchronously distributing workloads across multiple heterogeneous remote computing resources (e.g., local workstations, campus clusters, and commercial clouds). By automating job submission, monitoring, and data collection across distinct clusters, CLFF efficiently utilizes fragmented computational infrastructures, reduces queue times, and significantly accelerates the concurrent learning loop for large-scale projects.

\section{Software Description}
CLFF is implemented in Python and structured around a modular, object-oriented architecture based on two principal base classes. The \verb|Workflow| class provides an abstract foundation, defining a common interface for execution, stage management, and result handling. Workflows consist of an ordered sequence of computational stages, each encapsulating a logically coherent set of operations (e.g., model training, MD sampling, or DFT calculations). This stage-based design ensures transparent progress tracking and the seamless resumption of interrupted runs. The \verb|RemoteOperation| class abstracts remote computing interactions across heterogeneous environments, handling job submission, status monitoring, result retrieval, and error handling. Leveraging DPDispatcher \cite{yuan2025DPDispatcher}, it provides a unified interface to diverse connectivity protocols, enabling seamless workload distribution without requiring users to manage system-specific submission procedures. Together, these classes establish a versatile framework capable of implementing various multi-stage workflows with different complexity levels (e.g., MLIP concurrent learning, data generation, phonon calculation, potential energy surface scanning), while integrating smoothly with various computational backends.

CLFF provides a comprehensive command-line interface (CLI) alongside a Python API for advanced scripting and pipeline integration. The CLI features dedicated entry points for standard workflows (e.g., \verb|clff_cl|, \verb|clff_gendata|, \verb|clff_phonon|, \verb|clff_pes|). Command-line execution is straightforward:
\begin{Verbatim}
  clff_<workflow_name> PARAMETERS MACHINES
\end{Verbatim}
Here, the \verb|PARAMETERS| file defines workflow-specific settings (e.g., MLIP architecture, sampling strategies, and convergence criteria), while the \verb|MACHINES| file configures the computing resources (e.g., connection protocols, credentials, and schedulers). Both files utilize an intuitive, human-readable YAML format. This decoupled configuration allows researchers to maintain a reusable library of workflow templates while deploying them across diverse computing facilities simply by swapping the \verb|MACHINES| file.

Once launched, CLFF automates the entire workflow: data collection, script generation, model training, MD/DFT execution, uncertainty analysis, configuration selection, and iteration until convergence. The framework natively supports fault-tolerant execution, resuming from intermediate stages if interrupted. Real-time logging through an informative console (Figure S1) ensures transparent progress tracking and allows for manual intervention when necessary.

\subsection{Concurrent Learning Workflow}
\begin{figure}[!ht]
  \centering
  \includegraphics{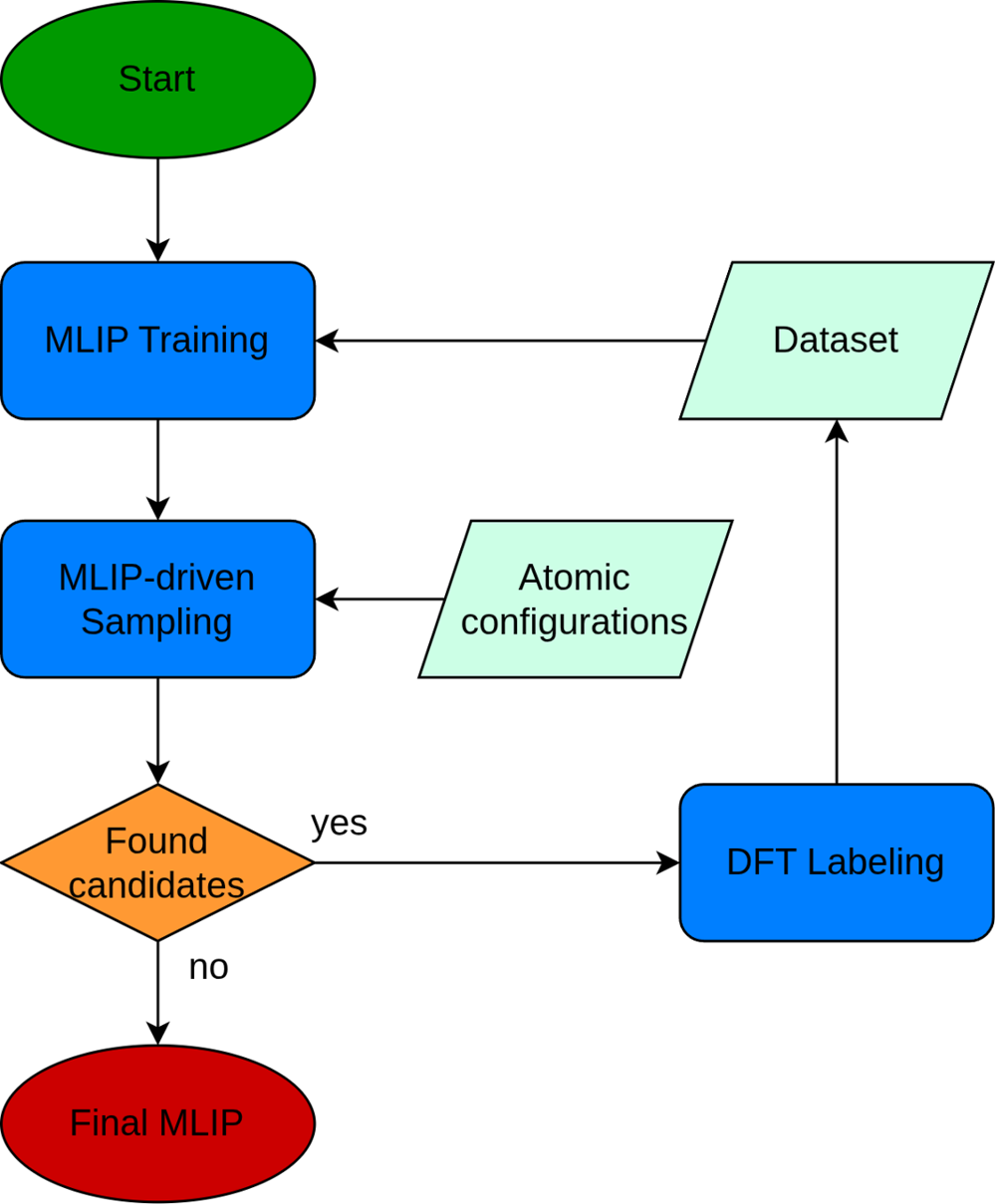} 
  \caption{The MLIP concurrent learning workflow implemented in CLFF. The distributed workflow management system (detailed in Section \ref{sect_workload_distribution}) is implemented across all blue blocks.}%
  \label{fig1_concurrentLearn}
\end{figure}

The concurrent learning workflow (\verb|clff.cl|) operates in iterative cycles following the algorithmic structure outlined in Figure \ref{fig1_concurrentLearn}. Each cycle trains the MLIPs on the current dataset, explores the configurational space via MLIP-driven MD simulations to identify highly informative configurations, and labels these selected configurations using high-fidelity DFT calculations to systematically augment the training set.

\subsubsection{MLIP Training}
CLFF natively supports training the SevenNet graph-based MLIP architecture \cite{park2024Scalable}. Through its integration with the Graph-pes package \cite{gardner2026graphpes}, it extends compatibility to various leading graph-based models, including MACE \cite{batatia2023MACE}, NequIP \cite{batzner2022NequIP}, TensorNet \cite{simeon2023TensorNet}, Orb \cite{rhodes2025Orbv3}, SchNet \cite{schutt2018SchNeta}, and PaiNN \cite{schutt2021Equivariant}, while its flexible interface supports custom architectures. Model uncertainty is quantified using a deep ensemble approach \cite{lakshminarayanan2017Simplea}: multiple models are trained with different random initializations, and their predictive variance serves as the uncertainty metric. High inter-model disagreement pinpoints regions where the ensemble lacks reliable data, providing a robust active learning criterion to guide configuration selection.

\subsubsection{MLIP-driven Sampling}
To thoroughly explore phase space, CLFF drives MD simulations using LAMMPS \cite{thompson2022LAMMPS} or ASE \cite{hjorthlarsen2017atomic}, integrated with PLUMED \cite{tribello2014PLUMED} for enhanced sampling across wide temperature and pressure ranges. It also incorporates Grimme-D3 vdW corrections \cite{grimme2010consistent, grimme2011Effect, smith2016Revised}, which are vital for simulating layered materials, molecular crystals, and organic molecules. Combining Grimme-D3 corrections with PLUMED is highly effective for exploring complex chemical spaces like organic solvent reactions, macromolecular structures, and catalytic interfaces. The ensemble standard deviation of MLIP predictions quantifies the uncertainty of each sampled frame.

\paragraph{Energy Deviation:}
Let $E^b$ be the per-atom potential energy predicted by model $b$ in the MLIP ensemble. The standard deviation for the potential energy is computed as:
\begin{equation}
  \sigma_{E} = \sqrt{\left\langle {\left( E^b - \langle E \rangle \right)}^2 \right\rangle}
\end{equation}

\paragraph{Force Deviation:}
Let $\mathbf{f}_{i}^b$ be the force vector predicted by model $b$ on atom $i$, the deviation magnitude is calculated as:
\begin{equation}
\sigma_{f, i} = \sqrt{ \left\langle \left\| \mathbf{f}_i^b - \langle\mathbf{f}_i\rangle \right\|^2 \right\rangle }
\end{equation}
If relative force normalization is enabled, the criterion is offset by the magnitude of the mean force vector to avoid disproportionately penalizing high-force regions:
\begin{equation}
\sigma_{f, i, \text{rel}} = \frac{\sigma_{f, i}}{\| \langle\mathbf{f}_i\rangle \| + \gamma_f}
\end{equation}
where $\gamma_f$ is a small relative regularization constant. The overall force deviation for a given configuration is determined by the maximum atomic force deviation within that structure:
\begin{equation}
  \sigma_f = \max_{i} \left( \sigma_{f, i} \right)
\end{equation}

\paragraph{Stress Deviation:}
For stress, the standard deviation is evaluated over the independent components of the 6-dimensional Voigt notation vector $\mathbf{s}$:
\begin{equation}
  \sigma_{s,\alpha} =  \sqrt{ \left\langle {\left( s_{\alpha}^b - \langle s_{\alpha} \rangle \right)}^2 \right\rangle }
\end{equation}
where $s_{\alpha}^b$ is the $\alpha$-th component of the stress tensor predicted by model $b$. If enabled, relative normalization scales these components by the mean stress magnitude:
\begin{equation}
\sigma_{s, \alpha, \text{rel}} = \frac{\sigma_{s, \alpha}}{\| \langle\mathbf{s}\rangle \| + \gamma_s}
\end{equation}

The overall stress deviation for the configuration is determined by the maximum component deviation:
\begin{equation}
\sigma_{s} = \max_{\alpha} \left( \sigma_{s, \alpha} \right)
\end{equation}

Here, $\langle \dots \rangle$ denotes the MLIP ensemble average.

\paragraph{Candidate Selection:}
Configurations with elevated uncertainty are highly informative and targeted for DFT labeling. To optimize computational resource allocation, lower ($\sigma^{lo}$) and upper ($\sigma^{hi}$) thresholds are imposed for each property deviation (energy, forces, stress) to categorize configurations into three groups:
\begin{itemize}
  \item \textit{Accurate}: All deviations fall below $\sigma^{lo}$, indicating well-represented, trusted structures that require no further labeling.
  \item \textit{Inaccurate}: Any deviation exceeds $\sigma^{hi}$; these are unphysical or highly unstable configurations discarded to save computational resources.
  \item \textit{Candidate}: Structures belonging to neither of the above groups, representing regions where the MLIP is uncertain but physically plausible. Only these candidates are selected for DFT labeling.
\end{itemize}

The classification criteria are mathematically expressed as follows:
\begin{align} \label{eq_candidate}
  \mathit{Accurate} &\iff \begin{cases}
    (\sigma_{E} < \sigma_{E}^{lo}) \land (\sigma_{f} < \sigma_{f}^{lo}) \land (\sigma_{s} < \sigma_{s}^{lo}) & \textit{if stress included} \\
    (\sigma_{E} < \sigma_{E}^{lo}) \land (\sigma_{f} < \sigma_{f}^{lo}) & \textit{if stress excluded}
  \end{cases} \\[10pt]
  \mathit{Inaccurate} &\iff \begin{cases}
    (\sigma_{E} > \sigma_{E}^{hi}) \lor (\sigma_{f} > \sigma_{f}^{hi}) \lor (\sigma_{s} > \sigma_{s}^{hi}) & \textit{if stress included} \\
    (\sigma_{E} > \sigma_{E}^{hi}) \lor (\sigma_{f} > \sigma_{f}^{hi}) & \textit{if stress excluded}
  \end{cases}\\[10pt]
  \mathit{Candidate} &\iff \neg (\mathit{Accurate} \lor \mathit{Inaccurate})
\end{align}

Energy and forces are checked by default, while stress is optionally included. Thresholds are application-dependent; for example, a typical crystal simulation might employ $\sigma_{E}^{lo} = 0.05$~eV/atom, $\sigma_{f}^{lo} = 0.05$~eV/\AA, $\sigma_{E}^{hi} = 0.2$~eV/atom, and $\sigma_{f}^{hi} = 0.2$~eV/\AA.

\subsubsection{DFT Labeling}
CLFF natively interfaces with GPAW \cite{mortensen2024GPAW} for high-throughput DFT labeling, supporting plane-wave and real-space grid modes on both CPU and GPU hardware. Users can fully customize DFT parameters, such as exchange-correlation functionals, energy cutoffs, convergence criteria, and vdW corrections, to ensure that the labeled dataset is of high quality and relevant to specific research goals. Additionally, CLFF supports any DFT calculator with an ASE interface, e.g., VASP \cite{kresse1996Efficiency}, ABINIT \cite{gonze2002Firstprinciples}, CP2K \cite{kuhne2020CP2K}, Quantum ESPRESSO \cite{carnimeo2023Quantum}, etc. This feature gives users the flexibility to choose their preferred calculators for labeling. For instance, one may select computationally cheaper alternatives, such as the semi-empirical quantum mechanics method xTB \cite{bannwarth2021Extended}, to reduce costs for budget-constrained projects. Configuring custom ASE-compatible calculators in CLFF is straightforward (see Supporting Information). Labeling results are automatically parsed to augment the training dataset for the subsequent cycle.

CLFF also incorporates specific workflow control utilities:
\begin{itemize}
  \item \textit{Gradient Update-Based Training}: Employs a fixed number of gradient updates instead of fixed epochs, allowing predictable training runtimes as the dataset grows.
  \item \textit{Early MD Termination}: Aborts faulty MD trajectories early if anomalous box expansion or contraction occurs, saving resources during early exploratory cycles.
  \item \textit{Task Filtering}: Skips completed tasks upon workflow resumption and deduplicates candidate configurations against existing training dataset to avoid redundant DFT runs.
  \item \textit{Space Evaluation}: Analyzes the sampled configurational space to identify regions requiring additional exploration.
\end{itemize}

\subsection{Data Generation Workflow}
\begin{figure}[!ht]
  \centering
  \includegraphics{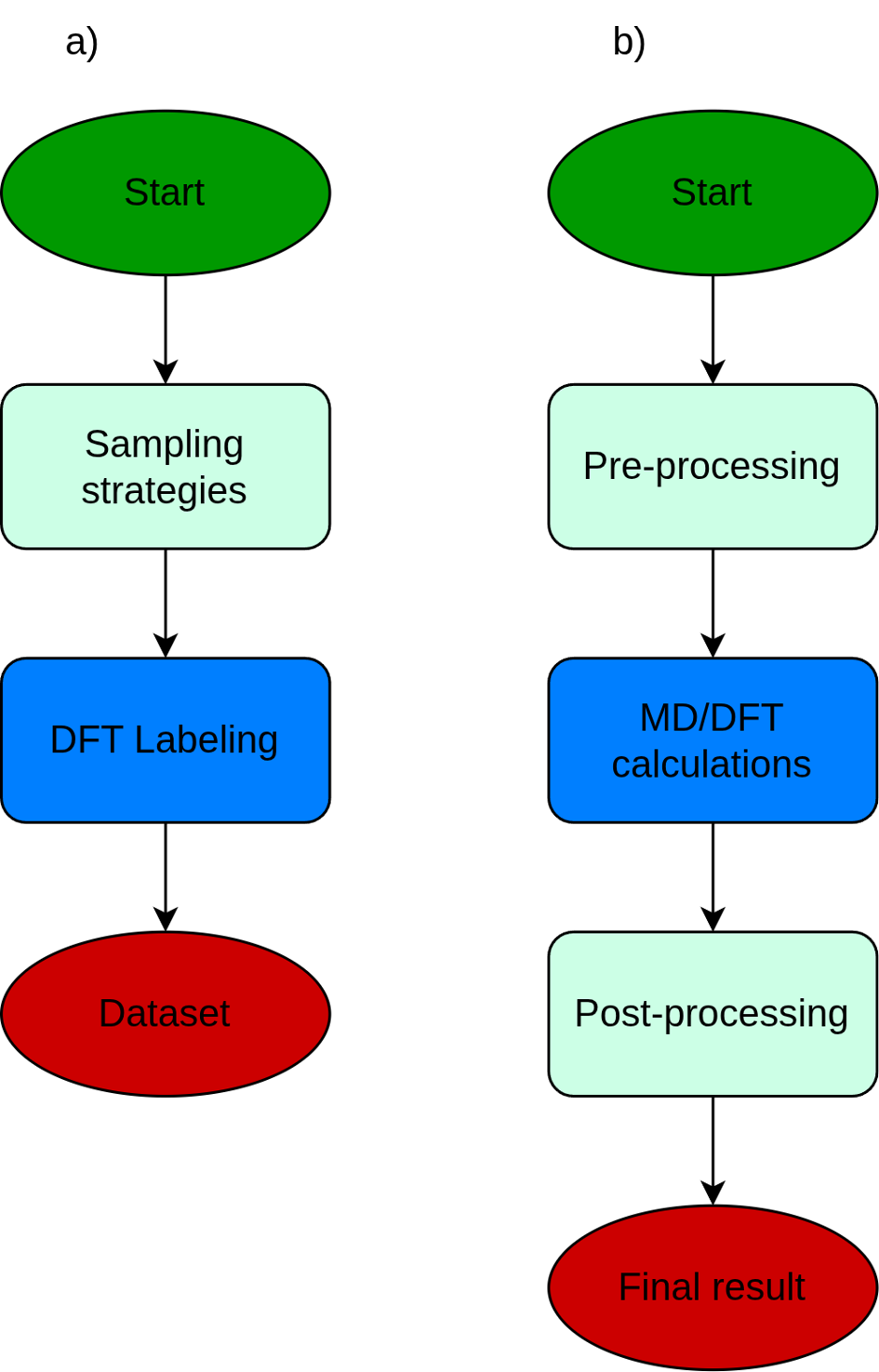}
  \caption{Workflows for data generation (a) and material properties calculations (b). The distributed workflow management system (detailed in Section \ref{sect_workload_distribution}) is implemented across all blue blocks.}%
  \label{fig2_gendata}
\end{figure}

Concurrent learning requires an initial baseline dataset to train the seed MLIP models. The data generation workflow (\verb|clff.gdata|; Figure \ref{fig2_gendata}a) is built for this purpose, though it can operate as a standalone utility independent of concurrent learning projects.

CLFF supports multiple native automated sampling strategies:
\begin{itemize}
  \item \textit{Random displacement sampling}: Perturbs equilibrium atomic positions to probe local potential energy surfaces and generate diverse baseline configurations.
  \item \textit{Ab initio molecular dynamics sampling}: Runs AIMD trajectories across diverse temperature and pressure ensembles to capture finite-temperature thermal fluctuations.
  \item \textit{Multidimensional strain sampling}: Applies systematic structural deformations and strain tensors to train models specifically on mechanical and elastic properties.
\end{itemize}

The framework accepts user-provided structures or automatically generates various types of systems (bulk crystals, 2D layers, surfaces, and molecules) leveraging ASE functionalities \cite{hjorthlarsen2017atomic} before labeling them via GPAW or ASE-compatible tools. Furthermore, CLFF includes robust \verb|extxyz| dataset manipulation utilities, supporting splitting, trajectory concatenation, key modification, custom frame filtering, deduplication, and structural sorting, ensuring clean, well-managed training dataset.

\subsection{Workflows for Material Property Calculations}
To facilitate downstream applications, CLFF implements subpackages under the \verb|clff.<props>| namespace to calculate material properties using MD, DFT, or trained MLIPs (Figure \ref{fig2_gendata}b). Supported modules include phonon dispersion calculations (integrated with Phonopy \cite{togo2023Implementation}) and potential energy surface (PES) scans. These highly customizable modules leverage backends like LAMMPS, GPAW, ASE-compatible calculators, and feature built-in post-processing utilities that automatically visualize and plot the computed materials properties.

\subsection{Workload Distribution}\label{sect_workload_distribution}
\begin{figure}[!ht]
  \centering
  \includegraphics{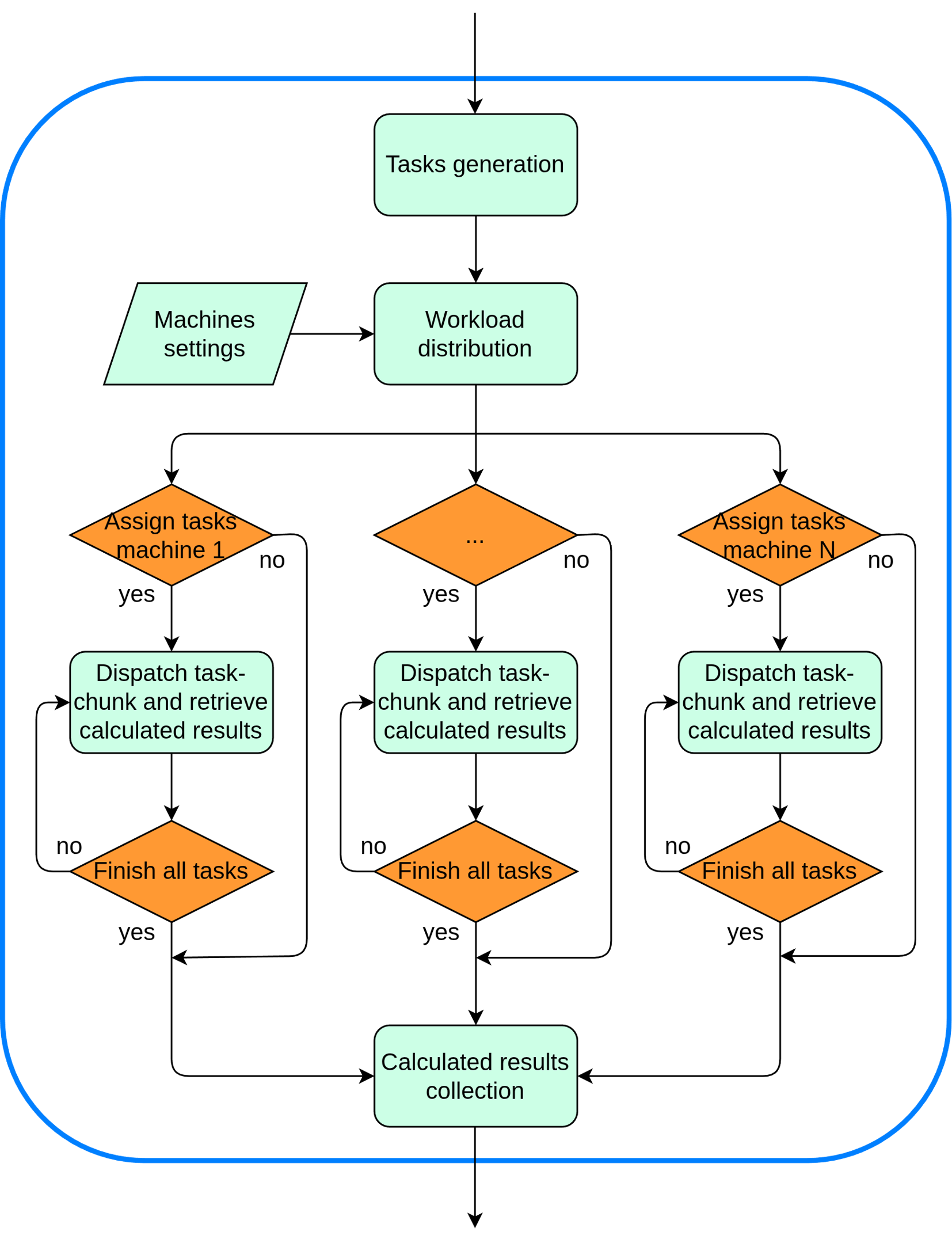}
  \caption{Schematic representation of the distributed workload management algorithm. This workflow is implemented across all blue blocks in the flowcharts of Figures \ref{fig1_concurrentLearn} and \ref{fig2_gendata}.}%
  \label{fig3_distribute_workload}
\end{figure}

One of the core features of CLFF is its ability to asynchronously distribute workloads across heterogeneous, remote computing clusters. Leveraging DPDispatcher functionalities \cite{yuan2025DPDispatcher}, CLFF concurrently manages multiple independent clusters with distinct schedulers (SLURM, SGE, TORQUE), connectivity protocols (SSH, OpenAPI), and hardware architectures. For every blue block illustrated in Figures \ref{fig1_concurrentLearn} and \ref{fig2_gendata}, CLFF automatically generates computational tasks, distributes them across designated remote machines, tracks execution, and retrieves results upon completion. The underlying algorithm is outlined in Figure \ref{fig3_distribute_workload}. Configuring distributed workloads requires nothing more than adding machine entries in the \verb|MACHINES| file, such as:
\begin{Verbatim}[frame=lines, commandchars=\\\{\}]
  gpaw_1:
    # \ldots other machine settings
    workload_ratio: 0.6

  gpaw_2:
    # \ldots other machine settings

  gpaw_3:
    # \ldots other machine settings
\end{Verbatim}
CLFF handles distribution automatically. Users can assign a specific \verb|workload_ratio| to each machine; if left unspecified, the remaining workload is distributed evenly. In this example, machine \verb|gpaw_1| is assigned 60\% of the workload, while machines \verb|gpaw_2| and \verb|gpaw_3| split the remaining 40\% equally (20\% each).

This workload distribution enables researchers to pool multi-institutional resources, from local workstations to campus HPC clusters and cloud infrastructure, or to balance workloads within a single cluster featuring heterogeneous nodes (e.g., routing massive supercells to high-memory nodes and smaller ones for standard nodes). Crucially, while modern DFT codes like GPAW \cite{mortensen2024GPAW} support both CPU and GPU architectures, they cannot combine them within a single execution block. CLFF circumvents this by splitting a batch of DFT calculations across CPU and GPU environments within the same workflow. By unifying submission, tracking, retrieval, and runtime estimation across disparate backends, CLFF effectively aggregates fragmented computing resources and makes them operate as a single computing center.

\subsection{Utilities}
CLFF includes a suite of supplementary utilities to streamline setup and enhance robustness, including but not limited to:
\begin{itemize}
  \item \textit{Automatic structure generation}: Interfaces with ASE \cite{hjorthlarsen2017atomic} to generate initial configurations from user-defined parameters, ensuring setup consistency.
  \item \textit{Flexible calculator integration}: Supports any ASE-compatible calculator sampling and labeling.
  \item \textit{Results visualization}: Automatically plots computed material properties for immediate post-processing insights.
  \item \textit{Parameter validation}: Pre-checks input files against expected schemas before workflow initialization.
  \item \textit{Fault-tolerant resumption}: Caches intermediate states to recover seamlessly from interruptions.
\end{itemize}

\section{Application}
To demonstrate the capabilities of CLFF, we present a representative application constructing the two-dimensional potential energy surface (PES) of a molybdenum disulfide (\ce{MoS2}) bilayer using both DFT calculations and MLIP-driven MD calculations.

\begin{figure}[!ht]
  \centering
  \begin{minipage}{0.48\textwidth}
    \centering
    \includegraphics{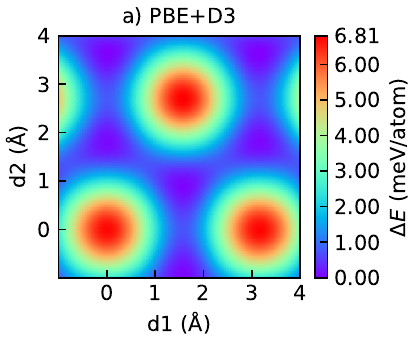}
  \end{minipage}
  \hfill
  \begin{minipage}{0.48\textwidth}
    \centering
    \includegraphics{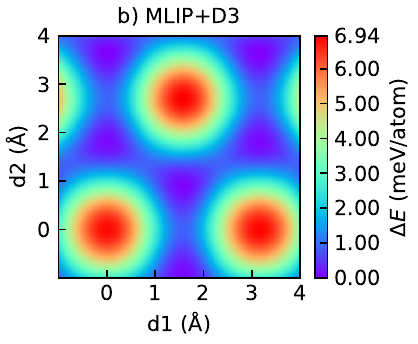}
  \end{minipage}
  \hfill
  \begin{minipage}{0.48\textwidth}
    \centering
    \includegraphics{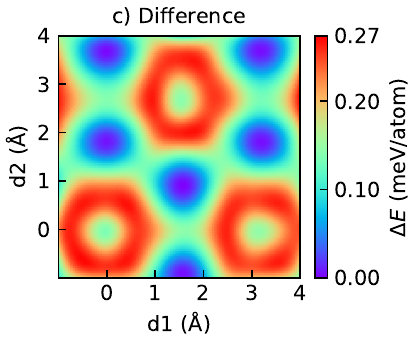}
  \end{minipage}
  \caption{PES of a bilayer 2H-\ce{MoS2} structure at a fixed interlayer distance of 6.64 \AA:\ (a) calculated using DFT, (b) calculated by MD simulation employing the SevenNet MLIP model, and (c) the difference between the two. The labels \textit{d1} and \textit{d2} represent the sampling grid shifts along the lattice vector directions. Subfigures (a) and (b) were automatically generated by the CLFF workflow.}%
  \label{fig5_pes_calculation}
\end{figure}

The PES landscapes for the bilayer 2H-\ce{MoS2} system from our calculations are illustrated in Figure \ref{fig5_pes_calculation}, demonstrating the exceptional quantitative accuracy achievable through automated data curation. Figure \ref{fig5_pes_calculation}a shows the DFT results, revealing distinct energy maxima (red regions) and minima (purple regions) corresponding to the AA-stacking and AB-stacking configurations, respectively. This confirms that the AB-stacking configuration is energetically more favorable than the AA-stacking configuration, in excellent agreement with established literature. The energy difference between these extrema is approximately 6.81 meV/atom, representing the energy barrier associated with interlayer sliding, which is a critical parameter for understanding the mechanical and tribological behavior of layered \ce{MoS2} structures.

Figure \ref{fig5_pes_calculation}b displays the PES landscape predicted by the SevenNet MLIP. The qualitative topography predicted by the MLIP matches the DFT results remarkably well, preserving the overall symmetry and shape of the PES landscape. The interlayer sliding energy barrier predicted by the MLIP is 6.94 meV/atom, yielding excellent quantitative agreement with the DFT reference. As quantified by the pointwise energy error map in Figure \ref{fig5_pes_calculation}c, the absolute discrepancy across the entire sampling grid remains strictly within the sub-meV/atom range, peaking at a mere 0.27 meV/atom. This extraordinary precision validates two core assertions: (1) the SevenNet MLIP architecture is highly capable of capturing complex, registry-dependent interlayer interactions, and (2) the CLFF framework can effectively identify the relevant configurations required to train such an MLIP model.

It is worth mentioning that developing interatomic potentials for layered materials is exceptionally challenging, as the relevant physics spans short-range directional covalent bonding within each layer and weak, but highly registry-dependent van der Waals interactions between layers. Conventional empirical IPs often struggle to capture this subtle interplay, leading to severe discrepancies in predicted properties. Indeed, systematic assessments even on simpler systems like graphene showed that traditional empirical IPs can fail severely when capturing vdW-governed sliding energetics \cite{qian2021comprehensive}; for chemically more complex materials like \ce{MoS2}, this challenge is amplified. Our results demonstrate that graph-based MLIPs like the SevenNet architecture, trained effectively via the CLFF framework, can accurately map these intricate interlayer landscapes, offering a highly promising alternative to conventional empirical frameworks for layered materials modeling.

Details of calculations and all scripts required to reproduce the presented calculations are provided in the Supporting Information, alongside template scripts demonstrating the concurrent learning workflow used to build the SevenNet model. Additional application examples are available on the CLFF documentation site \cite{nguyen2026clffdocumentation}.

\section{Conclusion}
In this work, we introduced CLFF (Concurrent Learning Force Field), an open-source, end-to-end automated framework designed to address the bottlenecks of dataset curation in MLIP development. By transitioning from static, human-guided sampling to an automated, dynamic concurrent learning paradigm, CLFF minimizes reliance on expert intuition and mitigates human bias, enabling the efficient generation of high-quality datasets tailored directly to target physical phenomena. The framework offers broad compatibility with contemporary graph-based MLIP architectures, integrates seamlessly with LAMMPS, ASE, and GPAW, supports various MD/DFT calculators via the ASE interface, and handles enhanced sampling methodologies through PLUMED. Beyond dataset curation, CLFF provides a comprehensive suite of workflows for generating training data and calculating key material properties, establishing it as a versatile, self-contained ecosystem for the materials science community.

A core, distinguishing innovation of CLFF is its ability to natively and asynchronously distribute computational workloads across heterogeneous, remote HPC infrastructures. While existing packages are often constrained by uniform computing environments or tied to specific institutional architectures, CLFF addresses the practical realities of modern computational research, where investigators frequently rely on fragmented infrastructures spanning local workstations, campus supercomputers, and commercial cloud platforms. By automating real-time submission, monitoring, and data retrieval across diverse clusters, CLFF maximizes resource utilization, minimizes queuing bottlenecks, and significantly accelerates the concurrent learning loop.

As a future roadmap, we plan to expand the capabilities of CLFF to support a broader range of MLIP architectures, including next-generation equivariant graph-based models like EquiformerV3 \cite{liao2026EquiformerV3}, as well as classic neural network and kernel-based frameworks. Furthermore, we intend to incorporate non-ensemble approaches \cite{zaghen2026Uncertaintyaware} for advanced uncertainty quantification.

\section{Data Availability}
CLFF is publicly available as open-source code on GitHub at https://github.com/thangckt/clff/ under the MIT license, accompanied by comprehensive documentation and tutorials available at https://thangckt.github.io/clff

\section*{Supporting Information}
The following supplementary materials are available free of charge:
\begin{itemize}
  \item \verb|_install/|: shell scripts for installation.
  \item \verb|example_pes_gpaw/|: scripts to reproduce PES calculation using GPAW.
  \item \verb|example_pes_lammps/|: scripts to reproduce PES calculation using LAMMPS.
  \item \verb|example_pes_ASEcalculator/|: scripts to perform PES calculation using an ASE-compatible calculator.
  \item \verb|example_concurrent_learning/|: setting templates to perform PES concurrent learning with the SevenNet MLIP model.
\end{itemize}

\section*{Acknowledgments}
The authors gratefully acknowledge the supercomputing resources provided by the UNIST Supercomputing Center.

\biboptions{sort&compress} 
\bibliographystyle{elsarticle-num} 
\bibliography{figure/reference}

\includepdf[pages=-, link]{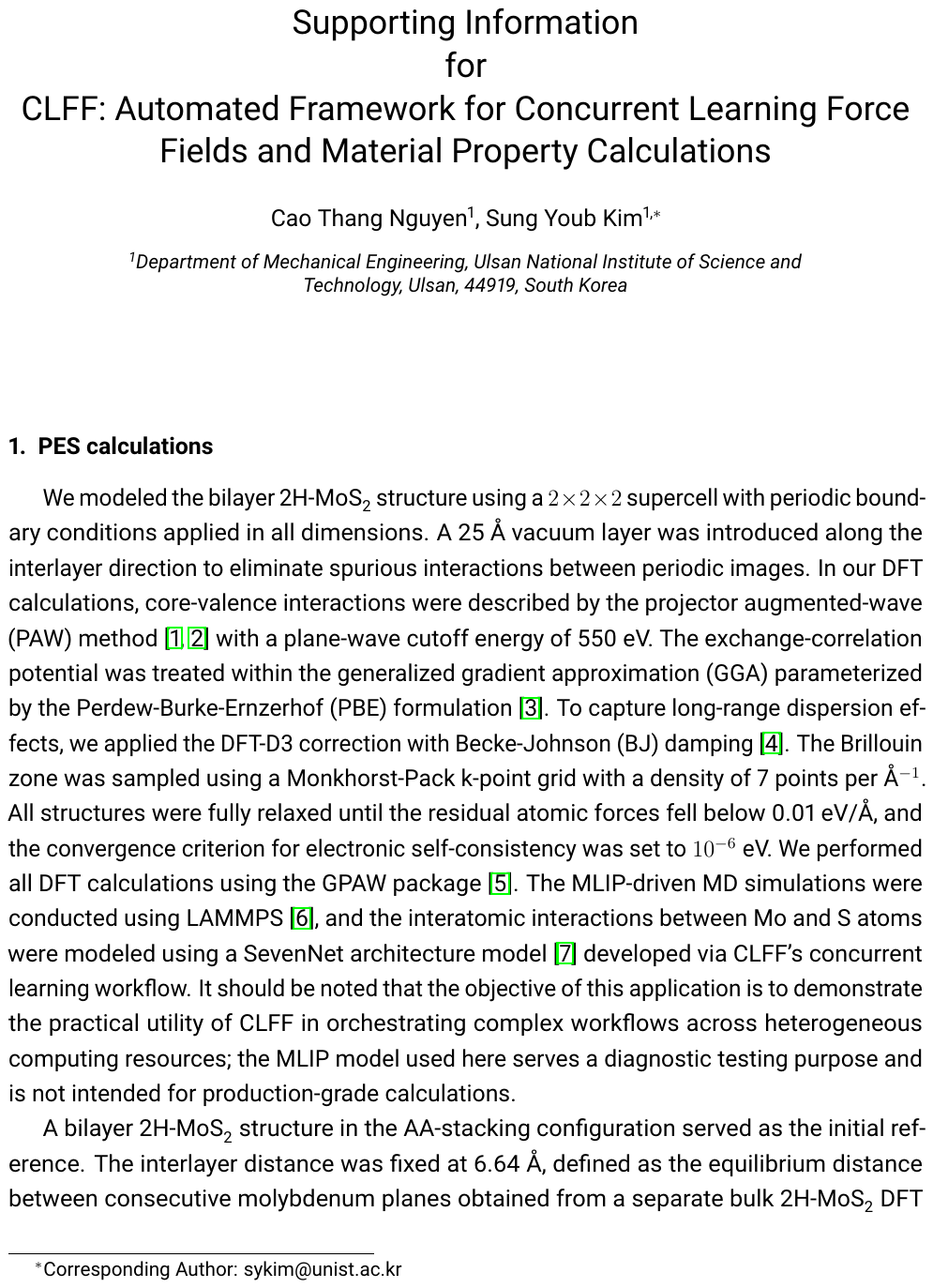}

\end{document}